\documentclass[]{spie}  %>>> use for US letter paper
\usepackage{amsmath,amsfonts,amssymb}
\usepackage{graphicx}
\usepackage[colorlinks=true, allcolors=blue]{hyperref}
\usepackage{array, makecell}

\title{Overview of focal plane wavefront sensors to correct for the Low Wind Effect on SUBARU/SCExAO}

\author[a,b,c]{S. Vievard}
\author[d]{S.P. Bos}
\author[e]{F. Cassaing}
\author[f]{A. Ceau}
\author[a,c,g,h]{O. Guyon}
\author[i]{N. Jovanovic}
\author[d]{C.U. Keller}
\author[a]{J. Lozi}
\author[f]{F. Martinache}
\author[f]{D. Mary}
\author[e]{A. Montmerle-Bonnefois}
\author[e]{L.M. Mugnier}
\author[f]{M. N'Diaye}
\author[j]{B. Norris}
\author[a]{A. Sahoo}
\author[g]{J-F Sauvage}
\author[d]{F. Snik}
\author[d]{M.J. Wilby}
\author[j]{A. Wong}

\affil[a]{National Astronomical Observatory of Japan, Subaru Telescope, 650 North A`ohoku Place,Hilo, HI 96720, U.S.A.}
\affil[b]{Observatoire de Paris - LESIA, 5 Place Jules Janssen, 92190 Meudon, France}
\affil[c]{Astrobiology Center of NINS, 2-21-1, Osawa, Mitaka, Tokyo, 181-8588, Japan}
\affil[d]{Leiden University, Rapenburg 70, 2311 EZ Leiden, Netherlands}
\affil[e]{ONERA, The French Aerospace Lab, University of Paris Saclay, F-92322 Châtillon - France}
\affil[f]{Observatoire de la Côte d'Azur, 96 Boulevard de l'Observatoire, 06300 Nice, France}
\affil[g]{College of Optical Sciences, University of Arizona, Tucson, AZ 85721, U.S.A.}
\affil[h]{Jet Propulsion Laboratory, 4800 Oak Grove Drive, MS 183-901, Pasadena, CA 91109, U.S.A.}
\affil[i]{California Institute of Technology, 1200 E California Blvd, Pasadena, CA 91125, U.S.A.}
\affil[j]{Sydney Astrophotonic Instrumentation Labs, University of Sydney, Australia}

\authorinfo{Further author information: Sebastien Vievard: E-mail: vievard@naoj.org}

\begin{document} 
\maketitle
\begin{abstract}
The Low Wind Effect (LWE) refers to a phenomenon that occurs when the wind speed inside a telescope dome drops below $3$m/s creating a temperature gradient near the telescope spider. This produces phase discontinuities in the pupil plane that are not detected by traditional Adaptive Optics (AO) systems such as the pyramid wavefront sensor or the Shack-Hartmann. Considering the pupil as divided in 4 quadrants by regular spiders, the phase discontinuities correspond to piston, tip and tilt aberrations in each quadrant of the pupil. Uncorrected, it strongly decreases the ability of high contrast imaging instruments utilizing coronagraphy to detect exoplanets at small angular separations.
Multiple focal plane wavefront sensors are currently being developed and tested on the Subaru Coronagraphic Extreme Adaptive Optics (SCExAO) instrument at Subaru Telescope: Among them, the Zernike Asymmetric Pupil (ZAP) wavefront sensor already showed on-sky that it could measure the LWE induced aberrations in focal plane images. The Fast and Furious algorithm, using previous deformable mirror commands as temporal phase diversity, showed in simulations its efficiency to improve the wavefront quality in the presence of LWE. A Neural Network algorithm trained with SCExAO telemetry showed promising PSF prediction on-sky. The Linearized Analytic Phase Diversity (LAPD) algorithm is a solution for multi-aperture cophasing and is studied to correct for the LWE aberrations by considering the Subaru Telescope as a 4 sub-aperture instrument.
We present the different algorithms, show the latest results and compare their implementation on SCExAO/SUBARU as real-time wavefront sensors for the LWE compensation.
 
\end{abstract}


\section{INTRODUCTION}
\label{sec:intro}  % \label{} allows reference to this section

Current science goals such as imaging/characterizing exoplanets or debris disks require both high angular resolution and high contrast imaging. The first requirement is fulfilled by the current large telescopes in the world, and will be even more satisfied with the next generation of Extremely Large Telescopes (ELTs). The second requirement is currently fulfilled by the combination of different instrumentation or data reduction techniques such as coronagraphs, Extreme Adaptive Optics (ExAO) and/or image post-processing. 
However, the performance of these techniques have two instrumental limitations. Firstly there are Non-Common Path Aberrations (NCPA) defined as the differential aberrations between the ExAO system and the science detector. Indeed, because the ExAO and the science optic paths are different, some aberrations on the science detector are not seen by the wavefront sensor (WFS). Secondly there is the Island Effect that is caused by the presence of the spiders. This Island effect can be divided in two categories of cause for the wavefront degradations: on one hand the spider-induced discontinuities of the pupil creates differential pistons not well seen by the WFS and therefore not well controlled (also named petalling or disconnectedness). On the other hand, the presence of the spiders creates, under some circumstances described hereafter, the Low Wind Effect (LWE). 
Both instrumental limitations constrain the achievable contrast, but we will focus here on the Low Wind Effect.

The LWE refers to a thermal effect. The temperature differential between the ambient air and the spiders ($T_{spider}<T_{air}$) induces radiative exchanges: the air gets cooler near the spider. This creates a refractive index gradient near the spiders that induces local delays of the wavefront. This phenomenon occurs especially when the wind speed is low (typically below $3$m/s), as the colder air is not blown away, and is to the best of our knowledge not detected by traditional WFS such as the Pyramid or the Shack-Hartmann.

%\begin{figure}[!h]
%    \centering
%    \includegraphics[width=0.4\linewidth]{Figures/LWE_explanation.png}
%    \caption{Low Wind effect explanation.}
%    \label{fig:lwe_intro}
%\end{figure}{}

\begin{figure}[!h]
    \centering
    \includegraphics[width=0.8\linewidth]{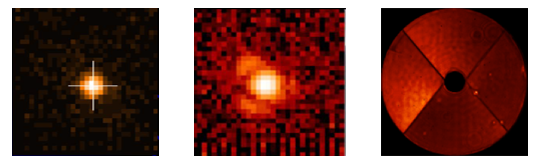}
    \caption{From left to right: PSF without LWE (DTTS) - PSF with LWE (DTTS) - Pupil plane instensity measured by ZELDA WFS on VLT/SPHERE. Source: \textit{Sauvage et al.}~\cite{sauvage2015low}}
    \label{fig:lwe_intro}
\end{figure}

While the impact of the LWE on the Adaptive Optics system itself is currently unknown, the impact on the Point Spread Function (PSF) of the telescope is clearly identified. Comparison of PSFs with and without LWE on VLT/SPHERE~\cite{beuzit2008sphere} acquired with the Differential Tip Tilt Sensor (DTTS, by \textit{Sauvage et al.}~\cite{sauvage2015low} - see Fig.~\ref{fig:lwe_intro}) showed that bright secondary lobes appear on the PSF in presence of the LWE (causing the degradation of high contrast capabilities for coronagraphs). 

{}

\textit{Sauvage et al.}~\cite{sauvage2015low} also performed a diagnostic of the pupil plane phase in presence of LWE using the ZELDA WFS on VLT/SPHERE. Allowing to convert the pupil phase in an intensity map as shown on Fig.~\ref{fig:lwe_intro} (right image) ZELDA showed that the effect of the LWE plus its impact on the Adaptive Optics system creates differential Piston, Tip and Tilt errors in each quadrant of the pupil.

In a very complete study of the Low Wind Effect, \textit{Milli et al.}~\cite{milli2018low} presented a corrective solution. He showed that radiative transfers between the spiders and ambient air can be strongly reduced by a new coating on the spiders. This was implemented on the VLT and was very efficient in reducing the LWE. From another point of view, solutions requiring new software/hardware developments should also be investigated.

When the modification of the telescope spider coating is not possible, or if the hardware of an instrument cannot be freely/easily modified, focal plane wavefront sensing seems to be the best solution to correct for the LWE since near-focal images of any source taken by a camera show the wavefront degradation as we see Fig.~\ref{fig:lwe_intro} (central image). The focal plane WFSs usually require minimal hardware modification since generally use existing detectors in the focal plane, e.g. the science detector. Another benefit is that they can simultaneously measure and correct NCPAs, another source of wavefront degradations.

As we explain in the first part of this paper, the instrument SCExAO on the Subaru Telescope is a unique platform to test new instrument concepts and algorithms. Focal plane wavefront sensors were or are being implemented on SCExAO to correct for the Low Wind Effect. The following of the paper is a presentation of four different focal plane wavefront sensors, their implementation requirements on SCExAO and the latest results obtained in simulation, in the lab on an internal source, or on-sky.

%\newpage
\section{THE LOW WIND EFFECT ON SCExAO}
\begin{figure}[!h]
    \centering
    \includegraphics[width=0.7\linewidth]{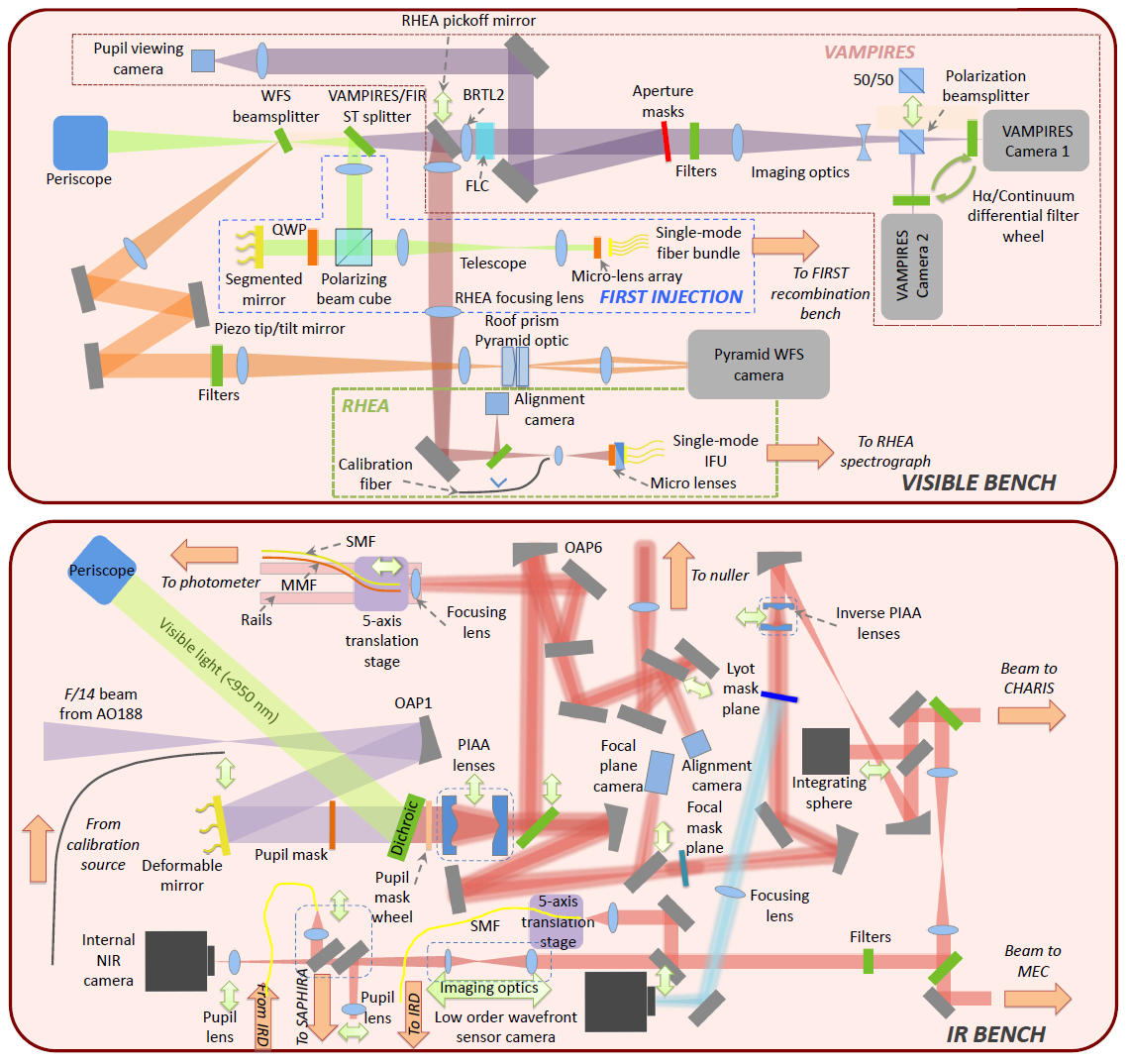}
    \caption{SCExAO IR and Visible bench.}
    \label{fig:scexao_visi}
\end{figure}

The Subaru Coronographic Extreme Adaptive Optics~\cite{2015PASP..127..890J} (SCExAO) instrument is located on the Infra-Red (IR) Nasmyth platform of the Subaru telescope. It is fed by AO188~\cite{minowa2010performance}, a 188-element Curvature sensor adaptive optics system that delivers a first stage of wavefront correction. SCExAO main WFS is a pyramid wavefront sensor (PyWFS) in the visible (around $800$nm - see \textit{Lozi et al.}~\cite{Lozi_2019} for more details) delivering a wavefront quality over 80\% Strehl (in H-band), critical for high contrast imaging with a coronagraph.
As shown on Fig.~\ref{fig:scexao_visi}, SCExAO hosts a lot of different modules and coronographs. SCExAO is divided over two benches:
\begin{itemize}
    \item The visible bench (from 600 to 950nm), hosting the PyWFS, VAMPIRES~\cite{norris2015vampires}, FIRST~\cite{huby2012first}, RHEA~\cite{rains2018development},
    \item the IR bench (from 950nm to around $2\mu m$) hosts the coronagraphs, CHARIS~\cite{Currie_2018}, MEC~\cite{walter2018mec}, GLINT~\cite{lagadec2018glint} and a fiber injection module for IRD~\cite{kotani2018infrared} .
\end{itemize}{}

Two of those modules, CHARIS and VAMPIRES, are currently open for science. The modularity of SCExAO makes it a unique platform to test new instruments concepts or new algorithms.

{}

SCExAO is victim of the LWE $10$ to $20\%$ of the time, leading to the impossibility to operate the instrument for science or some engineering tasks. Because of this, and thanks to its modularity, SCExAO is then a good platform to test new techniques for the LWE correction. The following sections present different algorithms that are (or are going to be) implemented on SCExAO to correct for the LWE.
%\newpage
\section{The Zernike Asymetric Pupil wavefront sensor}
\subsection{Principle}\label{zap_ppe}
\begin{figure}[!h]
    \centering
    \includegraphics[width=0.4\linewidth]{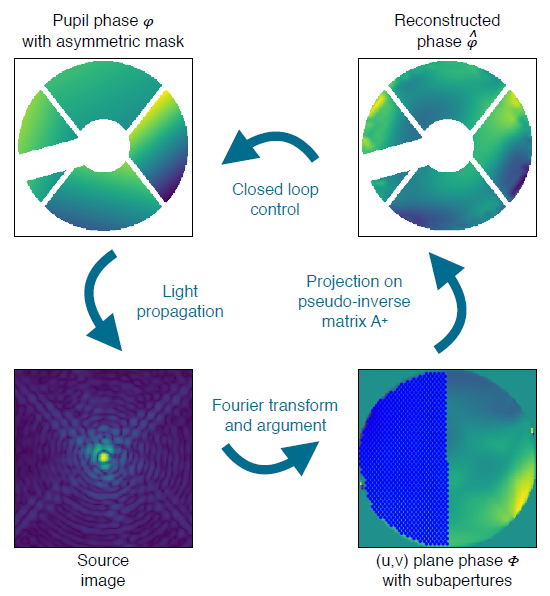}
    \caption{ZAP principle \cite{ndiaye_zap}}
    \label{fig:zap_principle}
\end{figure}

The first wavefront sensor we present is based on a Kernel phase analysis of the focal plane image of an unresolved source~\cite{martinache2013asymmetric} . In the small aberration regime ($\psi \ll 1$ rad), the relation between $\psi$, the phase in the pupil plane, and $\phi$ the phase in the Fourier space is linear:
\begin{equation}\label{ZAP_eq}
\phi = \mathbf{A}\psi,
\end{equation}
with $\mathbf{A}$ the transfer matrix between the two phases. This equation is easily inverted by computing $\mathbf{A^{\dag}}$ the pseudo-inverse of the rectangular matrix $\mathbf{A}$. The Singular Value Decomposition (SVD) of $\mathbf{A}$ gives: $\mathbf{A^{\dag}}=(\mathbf{A^{T}}\mathbf{A})^{-1}\mathbf{A^T}$. The inversion of Eq.~\ref{ZAP_eq} allows to compute the estimated phase aberration $\widehat{\psi}$ in the pupil plane from the phase $\phi$ measured in the Fourier space. However, $\widehat{\psi}$ suffers from a sign ambiguity. This degeneracy is lifted by introducing an asymmetry in the pupil plane, with a mask for example. Such a mask is shown Fig.~\ref{fig:zap_principle} top left. The process, shown on Fig.~\ref{fig:zap_principle} is therefore to compute the argument of the image Fourier transform and project the latest on the pseudo-inverse $\mathbf{A^{\dag}}$ to have an unambiguous estimation of the pupil phase $\widehat{\psi}$:
\begin{equation}\label{ZAP_res}
\widehat{\psi} = \mathbf{A}\phi.
\end{equation}
More details about the calibration of matrix \textbf{A} with Low Wind Effect modes from the instrument can be found in \textit{N'Diaye et al.}~\cite{ndiaye_zap}. 

{}

\subsection{Implementation on SCExAO}
\begin{figure}[!h]
    \centering
    \includegraphics[width=0.8\linewidth]{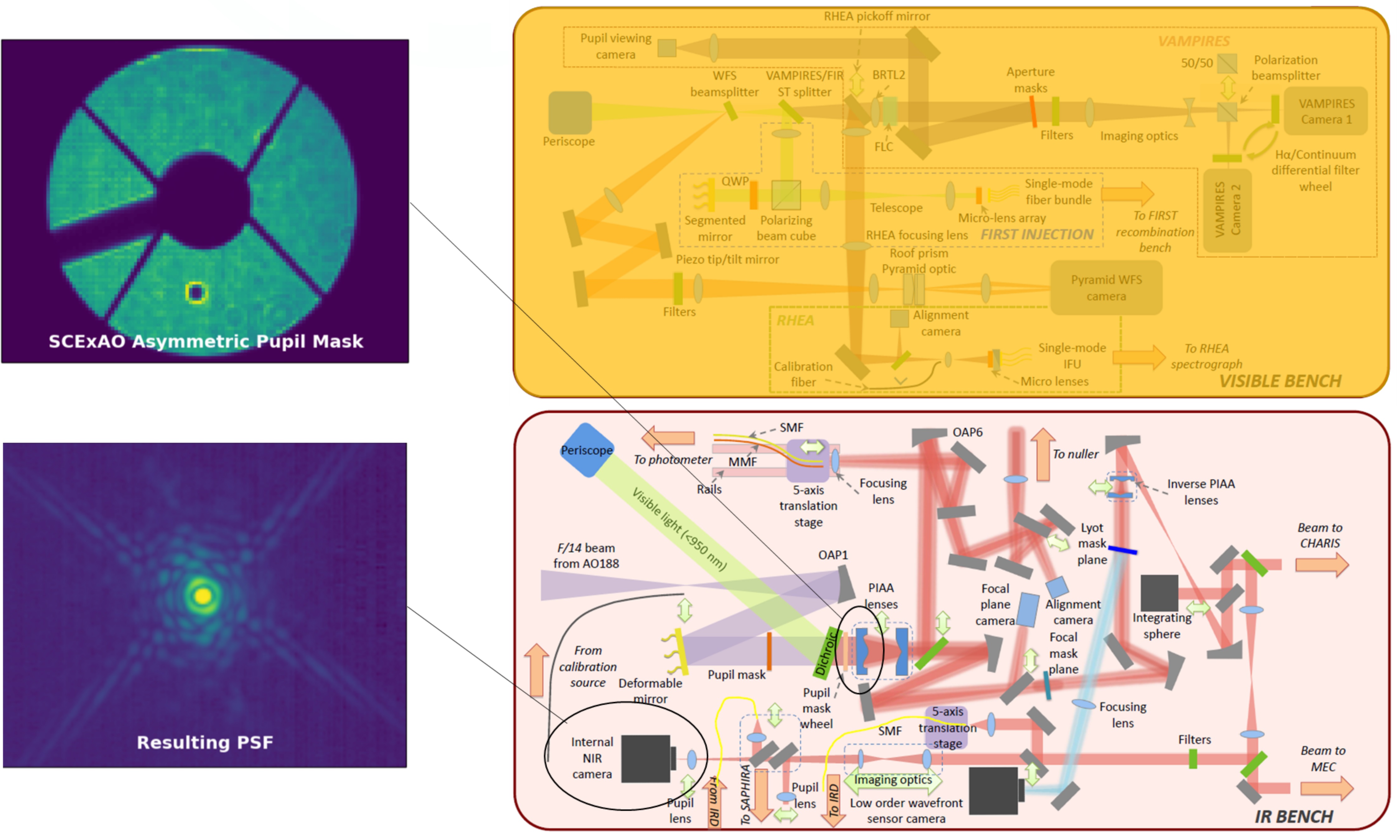}
    \caption{ZAP implementation in SCExAO: the asymmetric pupil mask is placed in a pupil plane filter wheel upstream the internal NIR camera. Images can then be recorded on the latest camera.}
    \label{fig:zap_implem}
\end{figure}

The requirements for the integration of the Zernike Asymetric Pupil wavefront sensor are :
\begin{itemize}
    \item a pupil plane mask,
    \item a focal plane image.
\end{itemize}
As we can see Fig.~\ref{fig:zap_implem} on the bottom right, SCExAO IR bench has a pupil mask wheel in the light path upstream the CRED 2 internal Near IR camera. The implementation is then straightforward and one of the pupil mask wheel slot contains the asymmetric mask shown Fig.~\ref{fig:zap_implem} top left. Images are then acquired with the internal NIR camera.

{}

\subsection{Results: in-lab and on-sky}

\begin{figure}[!h]\centering
\begin{tabular}{cccc}
	\includegraphics[width=0.2\linewidth]{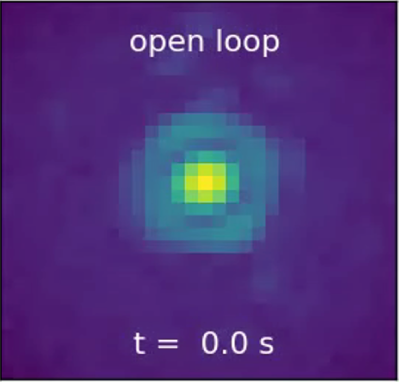} & \includegraphics[width=0.2\linewidth]{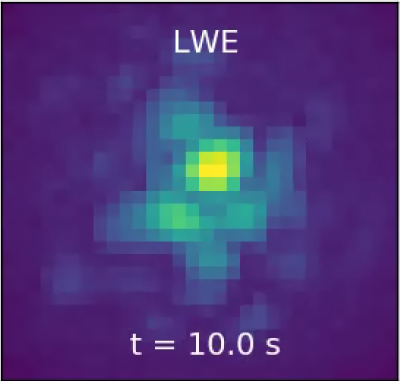} & \includegraphics[width=0.2\linewidth]{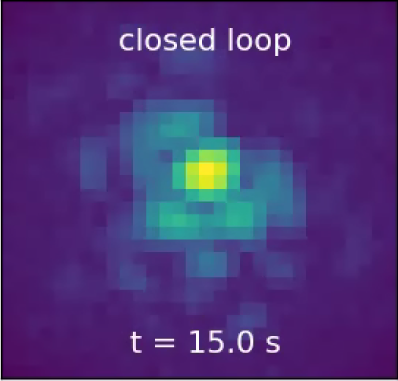} & \includegraphics[width=0.2\linewidth]{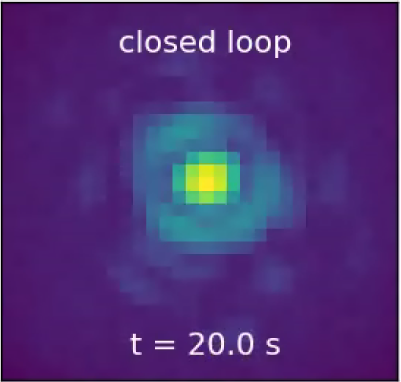}
\end{tabular}
	\caption{Closed-loop operation of ZAP algorithm on the SCExAO internal source~\cite{ndiaye_zap}. Images show from left to right the PSF without aberration (open loop), the PSF with aberrations (LWE) and the closed-loop operation at 2 different times.}
		\label{fig:closeloop-bench-ZAP}
\end{figure}

\textit{N’Diaye et al.}\cite{ndiaye_zap} paper shows results on SCExAO internal source and on-sky. The sensor response to each modes of the LWE was tested on H-band PSFs. Results on SCExAO internal source showed a linear response of the sensor over a range of about $100$ to $200$nm wavefront error. The algorithm was then tested in close loop, on the same internal source. A LWE aberration pattern was first applied on SCExAO deformable mirror. The second image from the left of Fig.~\ref{fig:closeloop-bench-ZAP} shows the aberrated shape of the PSF after applying the LWE aberration pattern (in comparison with the first image on the left). The algorithm then started the loop closure and stabilized the wavefront error to $20.3\pm2.8$nm RMS after around $20$ seconds (see Fig.~\ref{fig:closeloop-bench-ZAP} on the right), where the PSF is visually similar to its state before applying LWE aberrations.

\begin{figure}[!h]\centering
	\begin{tabular}{cc}
	\includegraphics[width=0.25\linewidth]{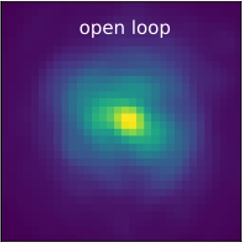} & \includegraphics[width=0.25\linewidth]{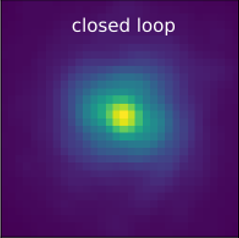}
\end{tabular}
\caption{Closed loop operation of ZAP on-sky~\cite{ndiaye_zap}. Focal image (left) in presence of LWE shows bright secondary lobes. Once the loop is closed with ZAP, the quality of the wavefront is improved, the bright lobes disapear (right image). A relative Strehl ratio of 37\% was measured between the two images.}
\label{fig:closeloop-sky-ZAP}
\end{figure}

Finally, the algorithm was tested on-sky in a closed-loop operation, in presence of LWE. Fig.~\ref{fig:closeloop-sky-ZAP} shows on the left the open loop, the PSF in presence of LWE acquired with the SCExAO/VAMPIRES module, in the visible (@$750$nm). We can see bright secondary lobes, sign of LWE. Image on the right shows the state of the PSF after closing the loop. An estimation of the Strehl ratio by measuring the relative intensity in those two images gives an improvement of $37\%$ before and after loop-closure.

\section{PSF reconstruction from PyWFS using Neural Network}
\subsection{Principle}
This method developed by \textit{Norris and Wong, 2019}, aims at training a deep neural network with synchronized in time PyWFS telemetry and focal plane images. Even though the training is slow and requires a large amount of data, its application can be done in real-time on modern GPUs equipped with tensor cores.

\subsection{Implementation on SCExAO}
The requirements for the integration of the Neural Network are :
\begin{itemize}
    \item a focal plane image,
    \item the telemetry delivered by the PyWFS.
\end{itemize}

\begin{figure}[!h]
    \centering
    \includegraphics[width=0.8\linewidth]{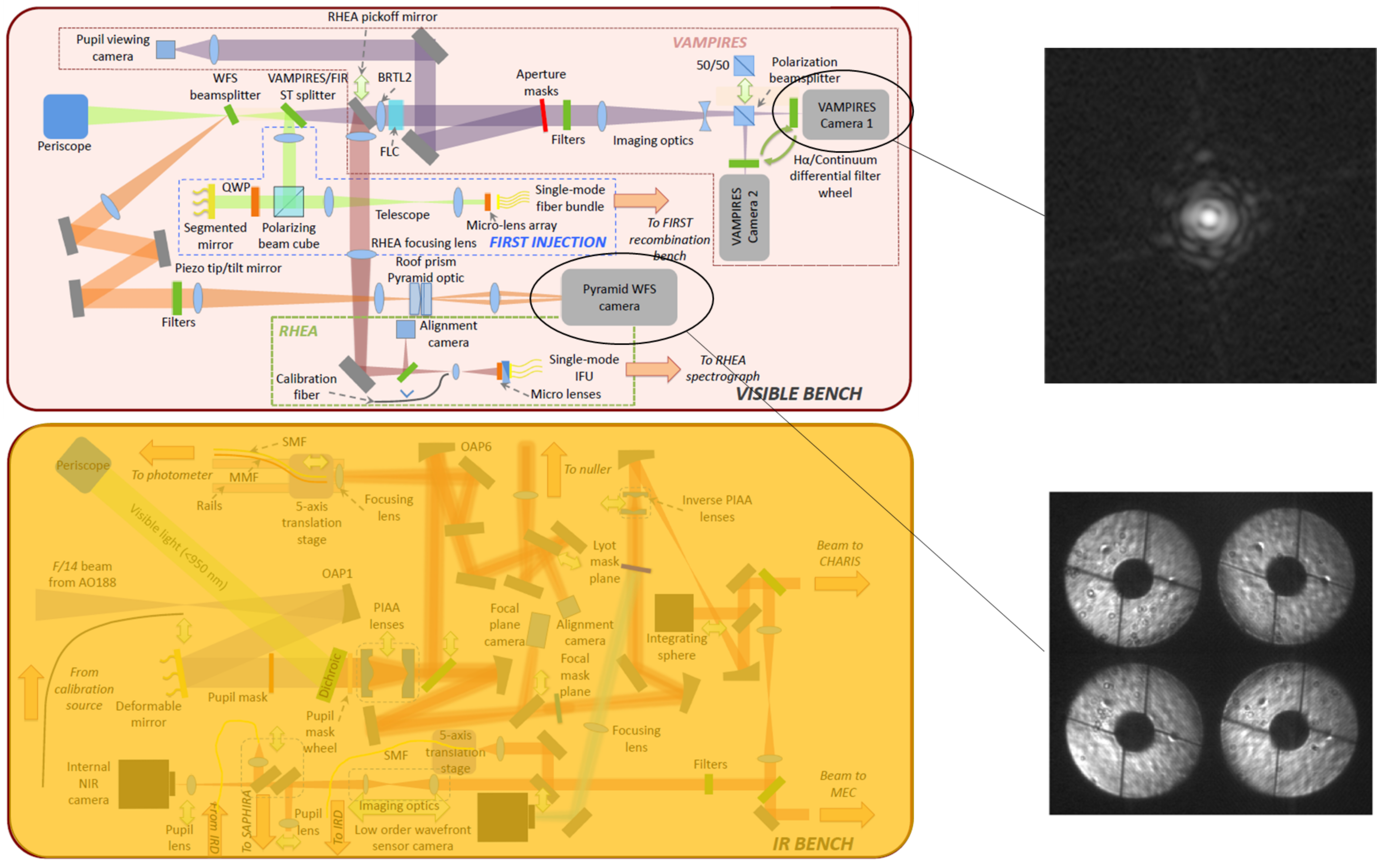}
    \caption{NN implementation in SCExAO: training of the NN needs the telemetry of the PyWFS and the focal plane image from one of VAMPIRES camera.}
    \label{fig:nn_implem}
\end{figure}{}

We can see Fig.~\ref{fig:nn_implem} on the left that the Neural Network is fed by one of VAMPIRES visible camera and the existing telemetry of the Pyramid. No extra hardware had to be added to implement the NEural Network.

\subsection{Results: on-sky}
\begin{figure}[!h]
    \centering
    \includegraphics[width=0.95\linewidth]{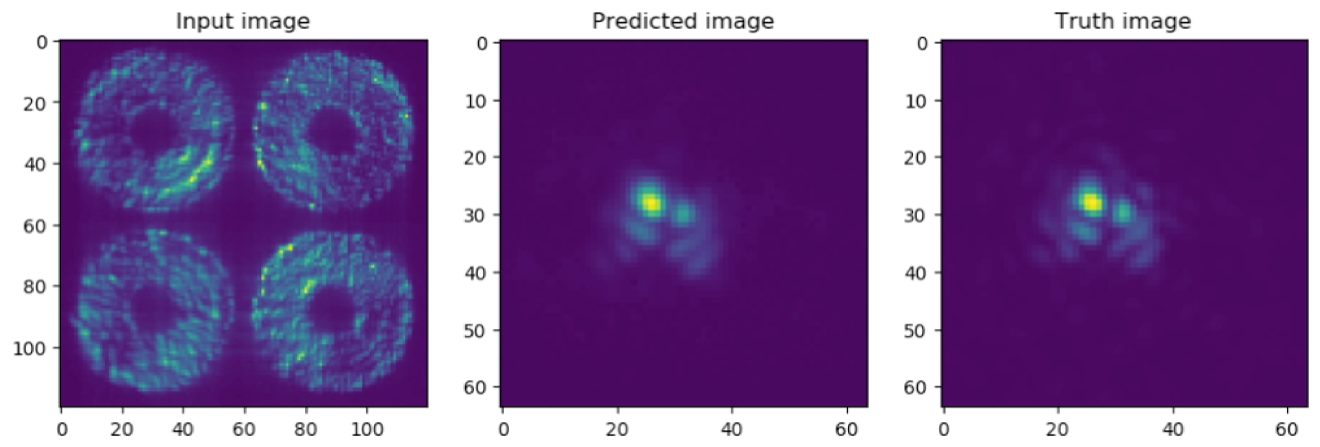}
    \caption{Neural Network on-sky results. Left: Input image for the NN, the PyWFS telemetry. Middle: Predicted image. Right: Real image. Courtesy: \textit{Barnaby Norris, University of Sydney}.}
    \label{fig:nn_onsky}
\end{figure}

Fig.~\ref{fig:nn_onsky} shows the on-sky demonstration of PSF prediction from PyWFS using telemetry. The left image is the PyWFS telemetry, the central image is the predicted image with the Neural Network and the image on the right is the actual true PSF image. The two images that are parts of a movie of several seconds are visually very similar. It is to be noted that the visible PSF reconstruction is highly non-linear and particularely sensitive to small wavefront errors.

{}
%\newpage

\section{The Linearized Analytic Phase Diversity}
\subsection{Principle}\label{lapd_ppe}
\begin{figure}[!h]
    \centering
    \includegraphics[width=0.4\linewidth]{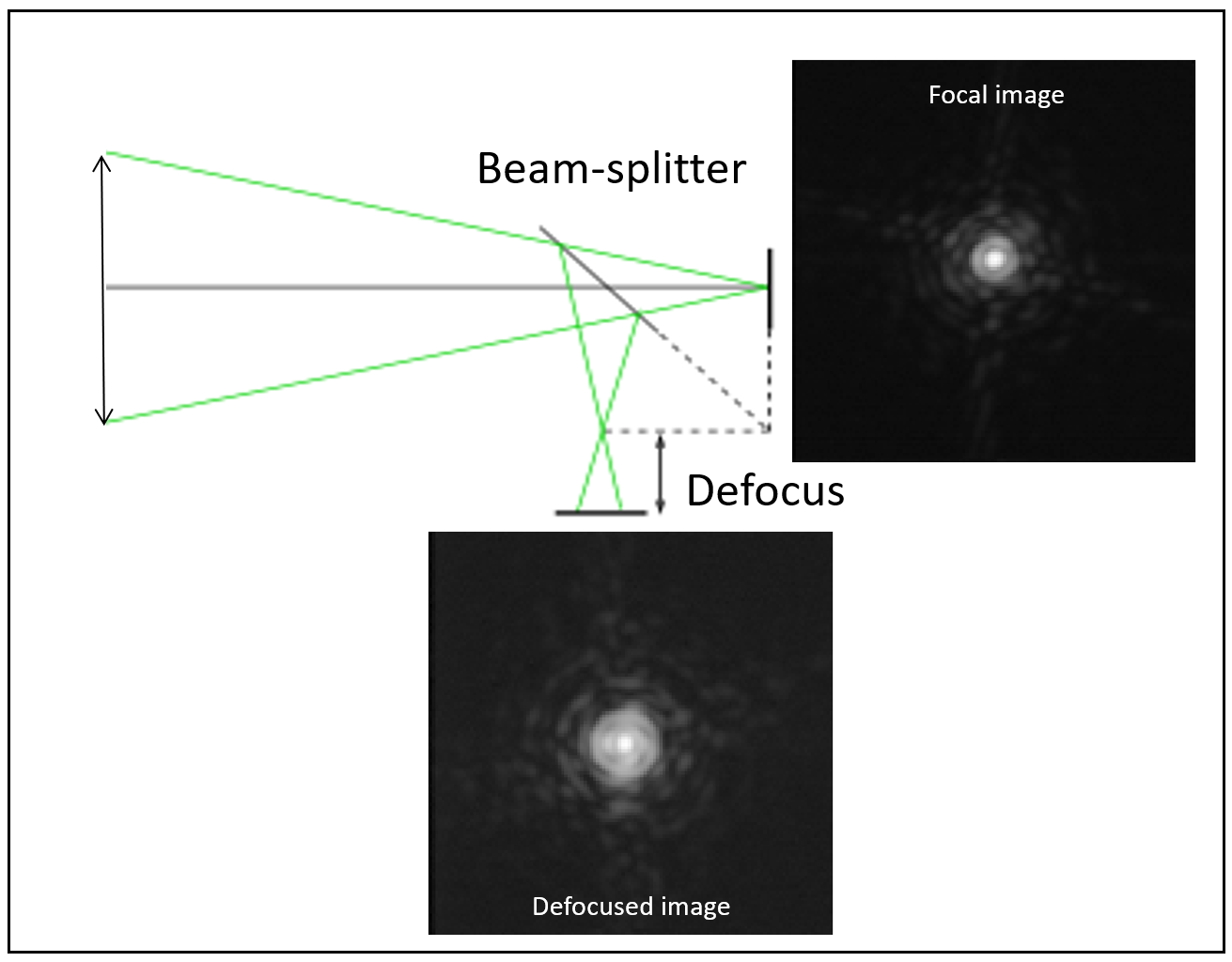}
    \caption{Phase diversity principle: simultaneous acquisition of focal and defocused images.}
    \label{fig:lapd_ppe}
\end{figure}

As its name indicates, the Linearized Analytic Phase Diversity (LAPD) algorithm is based on the phase diversity technique, in the case of small aberrations. Phase diversity technique consists in the acquisition of two images: a focal plane image and an image with a known aberration (here, a defocus - see Fig.~\ref{fig:lapd_ppe}).

{}
The acquisition of the second image allows to lift the degeneracy of the phase estimation based on only the focal plane image (see section~\ref{zap_ppe}). The two images are used to compute a criteria that is defined as the distance between the images and a perfect model of the telescope images. The aim is then to minimize this criteria. Since the link between the aberrations and the images is non-linear, it usually requires time-consuming, high computational cost iterative minimization algorithms.

In the context of small piston/tip/tilt estimation for multi-aperture telescopes cophasing, \textit{I. Mocoeur et al.}~\cite{Mocoeur-a-09b} computed a $1^{st}$ order Taylor expansion of the PSF. Using the latest to compute the phase diversity criterion led to an analytic solution for the aberration (more details are available in \textit{I. Mocoeur et al.}~\cite{Mocoeur-a-09b} letter). \textit{Vievard et al.}~\cite{vievard2018real} computed this algorithm with adding a small amount of iterations (typically less than 5), in order to increase its range and accuracy. LAPD was then validated experimentally. 

The idea here is to consider the Subaru telescope as a 4 sub-aperture telescope. Each of the LWE modes can be decomposed as piston/tip/tilt in each quadrant, acting like a multi-aperture mirror. LAPD is also able to estimate higher order aberrations on the full pupil.

\subsection{Implementation on SCExAO}
\begin{figure}[!h]
    \centering
    \includegraphics[width=0.8\linewidth]{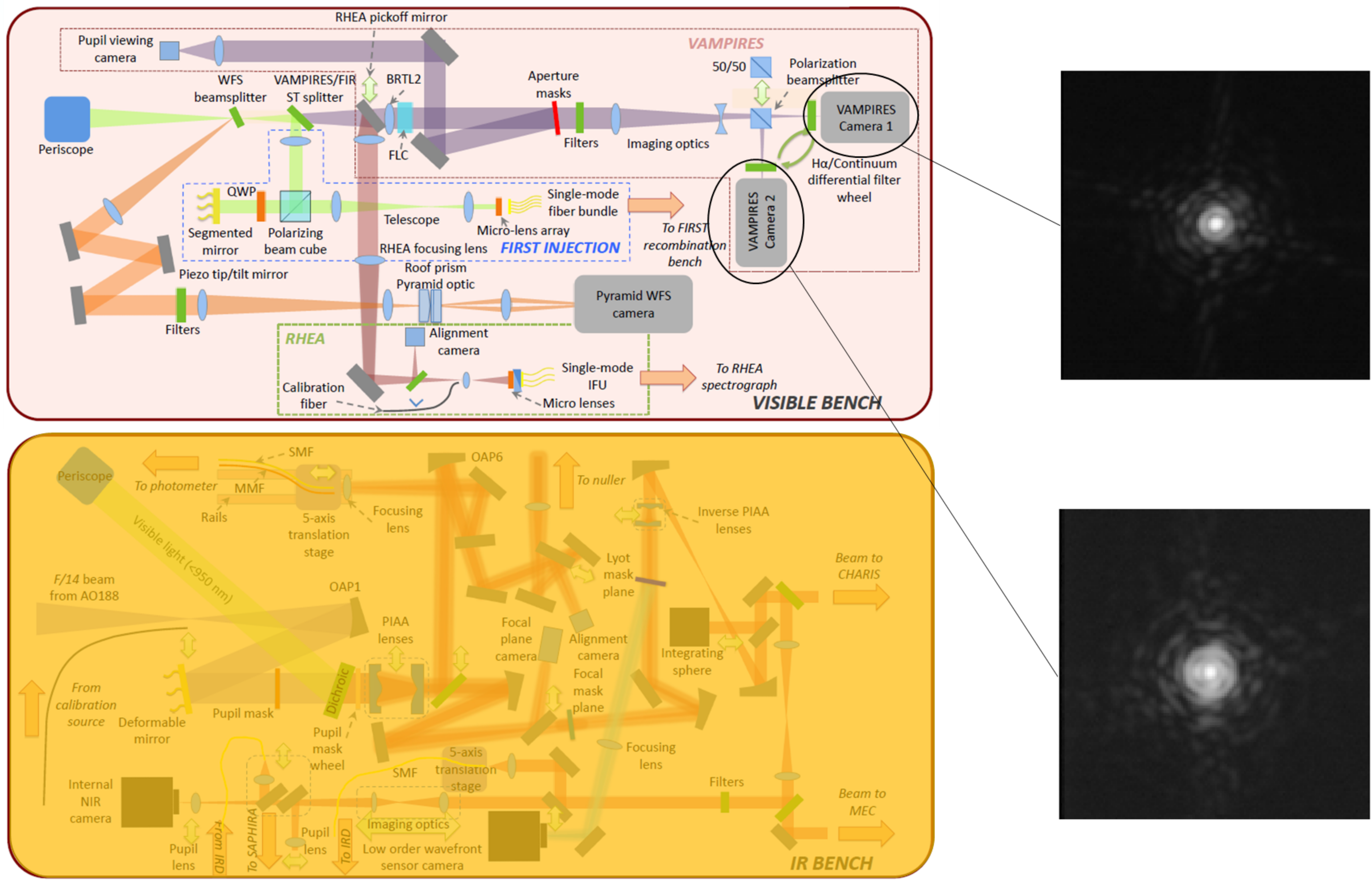}
    \caption{LAPD implementation in SCExAO: Two images can simultaneously be acquired with two cameras in the VAMPIRES module. Both cameras make focal plane cameras, but one sits on a translation stage allowing to introduce the desired phase diversity (defocus).}
    \label{fig:lapd_implem}
\end{figure}

The requirements for the integration of the LAPD wavefront sensor are :
\begin{itemize}
    \item a focal plane image,
    \item a defocused plane image.
\end{itemize}
We can see Fig.~\ref{fig:lapd_implem} top right that the two cameras of VAMPIRES instrument can simultaneously acquire an image. One of the camera being on a translation stage, it is easy to defocus the latest and have the two required images for LAPD. 
One could also record two images on the internal NIR camera. This camera is also on a translation stage, and can acquire an image in the focal plane, translate the camera and acquire an image in the defocused plane. However, there can be two drawbacks to this. The first drawback would be that this implementation would not be fast enough to allow a real time correction. Secondly, the time between the two image acquisitions might not "freeze" the LWE and the differential aberration in the two images would not allow an estimation with LAPD. 
Lastly, one could imagine applying the defocus directly on the DM, and acquire the images on the internal NIR camera. This process could be fast enough to correct for the LWE with LAPD in a close loop control, but it would not allow to do science simultaneously with CHARIS or VAMPIRES.

{}

\subsection{Results: Simulations}
\begin{figure}[!h]\centering
		\begin{tabular}{ccc}
	\includegraphics[width=0.2\linewidth]{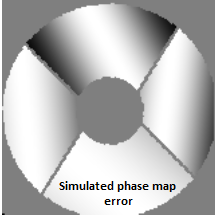} & \includegraphics[width=0.2\linewidth]{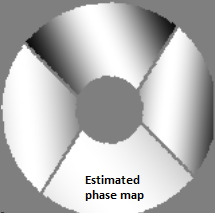} &
	\includegraphics[width=0.2\linewidth]{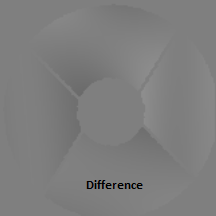}
\end{tabular}
\caption{Simulation of LWE modes on the Subaru pupil (left) - Estimation of the phase map by LAPD (center) - Difference between the two phase maps (right).}
\label{fig:estim-simu-lapd}
\end{figure}

Monochromatic images of an unresolved source were simulated with parameters that match the VAMPIRES cameras. Fig.~\ref{fig:estim-simu-lapd} left image shows the LWE modes, Piston/Tip/Tilt, simulated in each of the pupil quadrants. The RMS amplitude of the aberrations simulated here is of 1.3 rad. No high order aberrations were added for this simple simulation, and the images were generated with a high signal-to-noise ratio.
The amplitude of the defocus was set to $1$rad.

Fig.~\ref{fig:estim-simu-lapd} center image shows the phase estimation from LAPD. They visually look very similar, and it is even more obvious if we look at the difference between the input phase and the estimated phase (Fig.~\ref{fig:estim-simu-lapd} right image). We quantified the error on the estimation by the root mean square error of the estimation, which is $\lambda/100$ here. In this simple case, simulations confort us in LAPD ability to estimate LWE modes error for the Subaru pupil case.

\section{Fast \& Furious (sequential phase diversity)}

\subsection{Principle}

The last focal plane wavefront sensor that we present here is a sequential phase diversity wavefront sensor: Fast and Furious~\cite{keller2012extremely,korkiakoski2014fast} (F\&F). This algorithm was already studied in simulation and the lab for a SPHERE implementation to control the LWE~\cite{wilby2016fast,wilby2018laboratory}. On SCExAO, we aim for the on-sky validation of the algorithm. Instead of spatial phase diversity, as discussed in section~\ref{lapd_ppe}, this algorithm relies on a temporal phase diversity. The algorithm reconstructs the wavefront by splitting a single PSF image in its even and odd components. The odd component directly yields the odd wavefront error by simple algebra. For the even component one image is not sufficient to reconstruct the even wavefront error as the reconstruction suffers from a sign ambiguity. This degeneracy is lifted by a second PSF image with a known phase diversity, which is provided by the previous iteration and DM command of the F\&F loop. This has the major advantage that the wavefront sensor can be operated simultaneously with science observations as it continuously improves the wavefront. The algorithm is computationally very efficient as it requires only one Fourier transform and simple algebra per iteration.

\subsection{Implementation on SCExAO}
\begin{figure}[!h]
    \centering
    \includegraphics[width=0.8\linewidth]{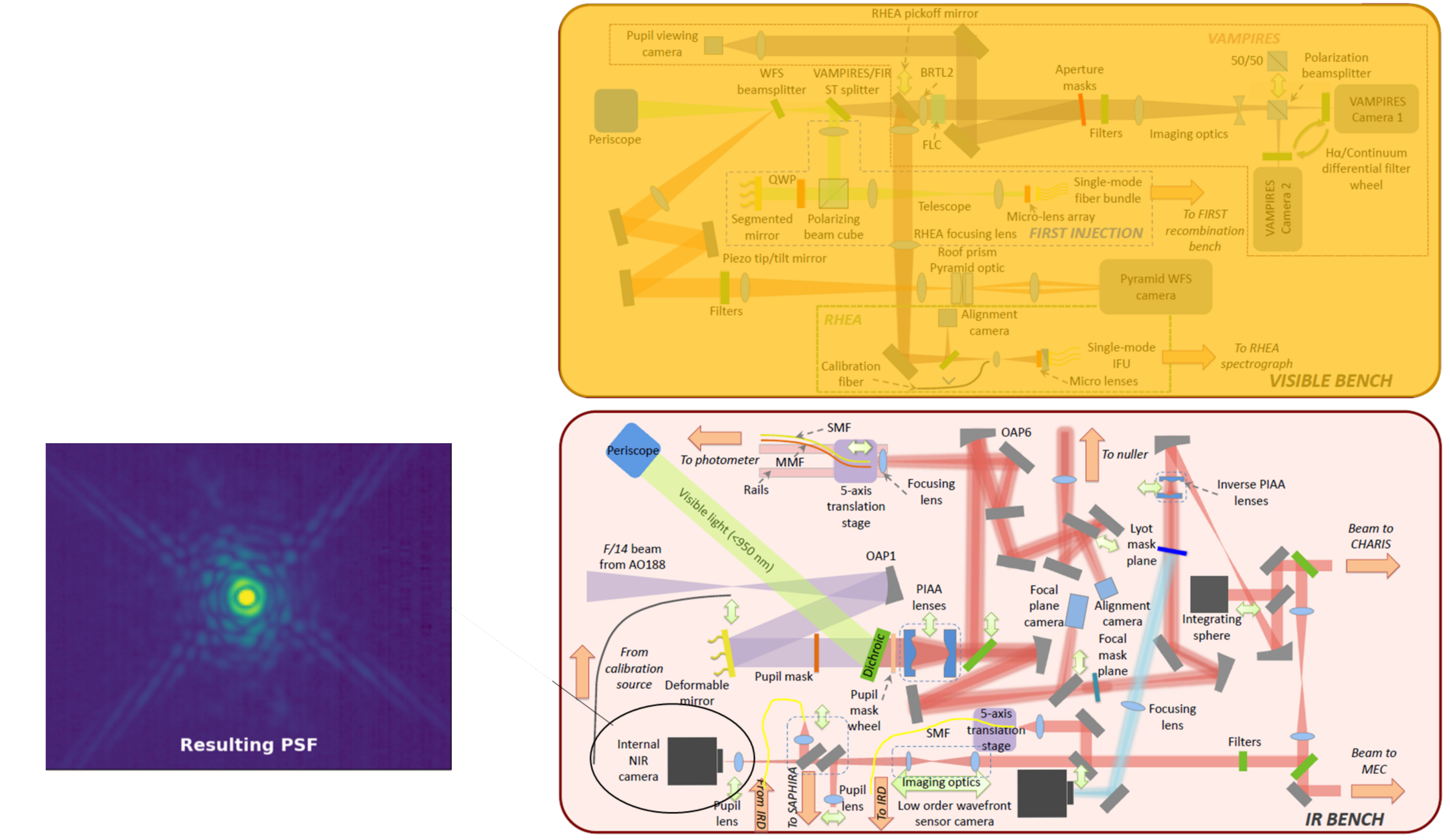}
    \caption{F\&F implementation in SCExAO only requires an image from the NIR camera and the DM stream.}
    \label{fig:ff_implem}
\end{figure}

There are two requirements for the integration of the F\&F WFS in SCExAO:
\begin{itemize}
    \item a focal plane image,
    \item an access to the DM.
\end{itemize}
For these simple requirements, we can just use the CRED 2 internal Near IR camera (see Fig.~\ref{fig:ff_implem}) and the DM stream. We can close the loop on the DM with F$\&$F while it is fed by the NIR camera images.

{}

\subsection{Results: simulations}
\begin{figure}[!h]\centering
	\begin{tabular}{cccc}
\includegraphics[width=0.2\linewidth]{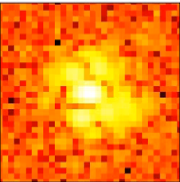} & \includegraphics[width=0.2\linewidth]{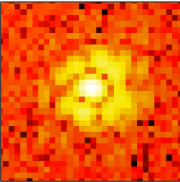} & \includegraphics[width=0.2\linewidth]{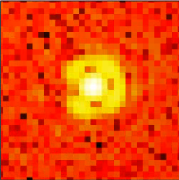} & \includegraphics[width=0.2\linewidth]{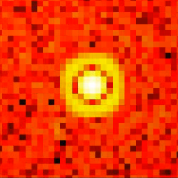}
\end{tabular}
\caption{Closed-loop operation of F\&F correcting for a static LWE in simulation: Focal plane images on various stages in the loop, going from uncorrected (left) to corrected (right). Courtesy: \textit{Steven P. Bos, University of Leiden}}
\label{fig:closeloop-simu-FF}
\end{figure}

Monochromatic images of an unresolved source were simulated with parameters that match the CRED2 NIR camera. Fig.~\ref{fig:closeloop-simu-FF} top row shows a closed-loop operation after simulating LWE modes on the PSF. We can see the first image on the left that present typical feature of LWE: bright secondary lobes. During the loop closure (from left to right) we can see that the PSF shape becomes more stable, until the last image where we can recognise the non-aberrated Airy pattern. 

\begin{figure}[!h]\centering
\includegraphics[width=0.4\linewidth]{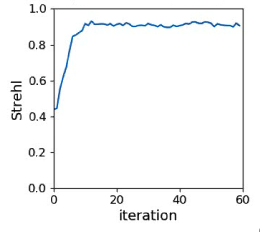}
\caption{Strehl ratio as function of iteration during the closed-loop simulation of F\&F. Courtesy: \textit{Steven P. Bos, University of Leiden}}
\label{fig:closeloop-simu-FF-Strehl}
\end{figure}

The graph on Fig.~\ref{fig:closeloop-simu-FF-Strehl} shows the evolution of the Strehl ratio during the loop closure. We can see that at iteration 0 the Strehl ratio is around 0.45, and that it goes up to 0.9 around iteration 15 and stays stable up to iteration 60. 
First tests on SCExAO internal source also showed promising results that still need to be quantified.

%\newpage

\newpage

\section{Conclusion}
\begin{figure}[!h]\centering
\includegraphics[width=\linewidth]{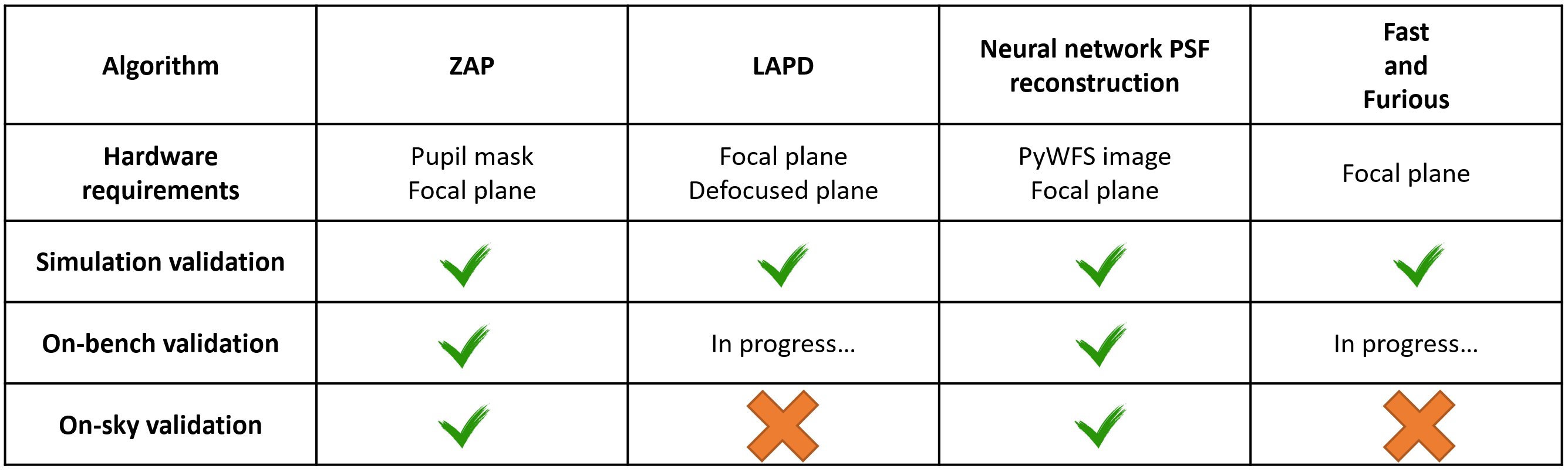}
\caption{Table of comparison between the different focal plane algorithms on SCExAO: Hardware requirements and progress in the integration on SCExAO.}
\label{fig:ccl}
\end{figure}

We presented four focal plane wavefront sensors implemented on SCExAO for the correction of the LWE. The appealing aspect of those WFS is that their integration in a system is very easy, but for some of them it is incompatible with acquisition of science data. Fig.~\ref{fig:ccl} shows a table that summarizes the hardware requirements for each of the different algorithms, and their integration/validation progress on SCExAO. We can see that two algorithm have already been validated on-sky, and two algorithms have been validated in simulation and are currently being integrated on SCExAO. In the nearby future, the idea would be to present an on-sky comparative performance of all these algorithms.

The modularity of SCExAO and the easy access to the hardware makes it a key platform to test and validate new concepts and algorithms. It is especially very useful for the next generation of ELTs, especially in terms of LWE since their gigantic dome might be subject to large air movements and temperature inhomogeneities.

\section{Acknowledgments}
The development of SCExAO was supported by the Japan Society for the Promotion of Science (Grant-in-Aid for Research \#23340051, \#26220704, \#23103002, \#19H00703 \& \#19H00695), the Astrobiology Center of the National Institutes of Natural Sciences, Japan, the Mt Cuba Foundation and the director's contingency fund at Subaru Telescope. F. Martinache's work is supported by the ERC award CoG - 683029. S.V. would also like to thank Julien Milli for the discussions on the Low Wind Effect. The authors wish to recognize and acknowledge the very significant cultural role and reverence that the summit of Maunakea has always had within the indigenous Hawaiian community. We are most fortunate to have the opportunity to conduct observations from this mountain.

% References
\bibliography{Main_lwe} % bibliography data in report.bib
\bibliographystyle{spiebib} % makes bibtex use spiebib.bst

\end{document}